\documentclass[manuscript]{acmart}
\AtBeginDocument{%
  \providecommand\BibTeX{{%
    \normalfont B\kern-0.5em{\scshape i\kern-0.25em b}\kern-0.8em\TeX}}}

\copyrightyear{2023}
\acmYear{2023}
\acmDOI{}

\begin{document}

\title[Collaborative Human-AI Design Thinking]{CHAI-DT: A Framework for Prompting Conversational Generative AI Agents to Actively Participate in Co-Creation}


\author{Brandon Harwood}
\affiliation{%
  \institution{IBM}
  \city{Dallas}
  \state{Texas}
  \country{USA}
}
\email{brandon.harwood@ibm.com}



\received{20 February 2023}
\received[accepted]{9 March 2023}

\begin{abstract}
    \textbf{Abstract: }This paper\footnote{This is a lightly revised and expanded-upon version of a paper originally presented at the ACM CHI 2023 Generative AI and HCI workshop.\cite{genai}} explores the potential for utilizing generative AI models in group-focused co-creative frameworks to enhance problem solving and ideation in business innovation and co-creation contexts, and proposes a novel prompting technique for conversational generative AI agents which employ methods inspired by traditional 'human-to-human' facilitation and instruction to enable active contribution to Design Thinking, a co-creative framework. Through experiments using this prompting technique, we gather evidence that conversational generative transformers (i.e. ChatGPT) have the capability to contribute context-specific, useful, and creative input into Design Thinking activities. We also discuss the potential benefits, limitations, and risks associated with using generative AI models in co-creative ideation and provide recommendations for future research.
\end{abstract}

\maketitle{

}
\section{Introduction}
In recent years, advancements in AI technology and accessibility have enabled the use of generative AI models such as ChatGPT \cite{openai-2023} as versatile tools for productivity, research, and creative application across a wide range of contexts and domains \cite{lin_2023} as well as the ability to successfully contribute to UX design functions. \cite{convo-designer-LLM} While these use-cases are often examined within the context of one-on-one human + AI interactions, little research explores how participants and facilitators might utilize generative creative agents effectively within group focused co-creative frameworks such as Design Thinking, which this paper posits could bring value to problem solving and ideation in business innovation and co-creation contexts. \cite{EDT-Kwon}

This paper seeks to explore the space of group + AI co-creative partnership, and proposes a prompting technique for conversational agents (i.e. ChatGPT) to co-create with humans, inspired by methods used in traditional 'human-to-human' facilitation and instruction typically seen in Design Thinking, a co-creative framework for multidisciplinary participants to collaborate and ideate together in groups to build user-focused solutions in business. \cite{IBM-approach-EDT} Through experiments using this prompting technique, I have gathered evidence that conversational generative transformers like ChatGPT have the capability to contribute context-specific, useful, and creative input into Design Thinking activities. I will also discuss the frameworks potential value in fostering ‘human/s + AI’ co-creative partnership within Kantosalo’s 5Cs Framework for Human-Computer Co-Creativity \cite{Kantosalo2020FiveCF} and IBM's Design Thinking Framework, described in \cite{IBM-approach-EDT}. \cite{EDT-Kwon}

Finally, I consider potential benefits, limitations, and risks associated with using generative AI in co-creative ideation, as well as speculate on opportunities for future research into how we might mitigate risks and measure effectiveness of ‘human/s + AI’ co-creative partnership. By addressing these limitations and exploring effective ways to partner with creative AI agents in group settings, I hope to set a path forward in advancing our understanding and use of AI systems/agents as helpful, effective partners and participants in co-creative pursuits.

\section{Background}
Human-computer co-creativity is described by Kantosalo et. al. as “the interactions within a human–computer \textit{collective}, the collective’s \textit{collaboration} process and creative \textit{contributions} to a \textit{community}, all situated within a rich \textit{context}.” \cite{Kantosalo2020FiveCF} This is contrasted with, but closely aligns to traditional human co-creative frameworks such as Design Thinking (DT), differentiated by the explicit account for the contribution of computational agents in co-creation. DT, as it is compared to Kantosalo’s model, can be described as employing a \textit{collective} group of individuals to utilize their “individual subjective” \textit{context} in reference to a \textit{community} (e.g. user or user-group, often included in DT workshops as participants) in a narrowly-scoped “idealized objective” \textit{context} defined in the setting of the DT session (e.g. problem space, process, etc.), so they may \textit{collaborate} with each other through the Design Thinking framework described in \cite{IBM-approach-EDT}, and ultimately \textit{contribute} creative artifacts (e.g. ideas, observations etc.). 

Kantosalo notes that “in a successful creative collaboration, the collective benefits from the profound communication and sharing of contributions within the collective.” \cite{Kantosalo2020FiveCF} This framework is differentiated from DT by specifying the “\textit{human-initiative vs. computational-initiative}” dichotomy, but this measure of success can also be used to describe a successful DT session, the only difference being that humans are the only participants with creative initiative in traditional DT. However, with recent research into the ability of generative AI agents to contribute to design spaces creatively through conversation \cite{convo-designer-LLM}, it is suggested that we are able to build a system respective of human vs. computational initiative, and integrate creative AI agents into group ideation spaces as active participants.

Drawing on this background, I intend to explore the potential of integrating creative AI agents into Design Thinking through a novel prompting framework, and in doing so, explore how we might be able to build environments which foster greater creative output in group-focused co-creation settings through mixed human-computer initiative.

\section{Methods}
\subsection{Initial Attempts and Findings}
OpenAI's ChatGPT \cite{openai-2023} was chosen as the testing grounds for experimentation due to it's availability and relatively advanced capabilities, as well as the conversational interface, which provide an environment enabling both DT facilitators and participants to engage with the model during live ideation. During initial testing it became apparent that a loose conversational style of interaction for 'facilitating' the model through DT exercises resulted in inconsistent output quality, requiring multiple back-and-forth explanations or re-framing of the prompts. This is sufficient (and sometimes ideal) for one-on-one ideation, but could be distracting in a live session. 

Through these loose interactions, however, it became clear that (much like human participants) the model performed the DT exercises with more success when provided a clear explanation of the activity, the background informing when the activity is appropriate to employ, step-by-step instruction, and context that informs why ideation is being performed to begin with. These patterns have helped inform two sets of instruction and explanation which, when provided to ChatGPT, result in better and more structured co-creative output from the model. Furthermore, when these instructions and explanations are provided at the beginning of the interaction with the model, it frees the facilitator to simply instruct the model to perform the exercise (or a step/subsequent steps of the exercise), and enables the facilitator/human participants to accept, reject, or add to the creative output of the model and continue their ideation.

\subsection{The CHAI-DT Framework}
The method I am introducing as the \textbf{Collaborative Human-AI Design Thinking (CHAI-DT) Prompting Method} is a convenient and adaptable framework combining these instructions and explanations into a single prompt that initiates the interaction and enables the conversational agent to contribute useful creative artifacts to groups in Design Thinking efforts. This is done by explaining the activity, defining success, and providing specific instructions and necessary context through a composition of six sub-prompts. The first four sub-prompts are ‘static instructions’ which introduce the structure and purpose of the chosen activity in the same way we might introduce it to human participants, and does not change between interactions with the agent. The final two are ‘variable instructions’, which are dependent on the session context and interaction needs from the facilitator and participants.

\textbf{Static Instructions:}
\begin{itemize}
\item {\verb|Introduction|}: “We are conducting a(n) '[Activity Name]' Design Thinking exercise.”
\item {\verb|Definition|}: Explanation of the exercise, it's intended outcome, and why it is useful for co-creation.
\item {\verb|Example(s)|}: One or multiple examples of an ideal end-state artifact.
\item {\verb|Instructions|}: Step-by Step instructions for the activity.
\end{itemize}
 
\textbf{Variable Instructions:}
\begin{itemize}
\item {\verb|Context|}: Appropriate context to inform the interaction. (i.e. Why are we co-creating? Who is participating? What background information can we provide?)
\item {\verb|Execute|}: Prompt the model to perform the activity, either in it's entirety or gradually (i.e. "perform the entire Hills exercise.", or "perform step 1.").
\end{itemize}

\section{Results}

\begin{table}
  \caption{ChatGPT's Responses to the 3 "Hills" DT Exercise Steps}
  \label{tab:freq}
  \begin{tabular}{p{1.4in} p{1.9in} p{2.1in}}\
        \textbf{"Who" (Potential Users)} & \textbf{"What" (Enablements)} & \textbf{"Wow" (Impact)} \\ \hline
        Retail store managers & Accurately predict sales trends & Dramatically reduce stockouts and overstocking, resulting in increased sales and profitability \\ \hline
        Inventory managers & Optimize inventory levels in real-time & Improve customer satisfaction by ensuring products are always in stock \\ \hline
        Supply chain managers & Identify underperforming products & Increase efficiency and productivity for store and inventory managers \\ \hline
        Sales associates & Identify overstocked products & Reduce waste and optimize resource utilization \\ \hline
        Customers (indirectly impacted by the system) & Determine reorder quantities and timing & Provide a competitive advantage in the retail industry through advanced data analytics and artificial intelligence \\ \hline
        Executives/decision-makers at RetailInc & Monitor stock levels and alert managers when stock falls below a certain threshold & Ensure accuracy and reliability of sales and inventory data, leading to better decision-making \\ \hline
        ~ & Provide insights into customer demand and behavior & Enhance the overall shopping experience for customers through better inventory management and product availability \\ \hline
        ~ & Generate automated reports and analytics for inventory and sales data & Enable RetailInc to respond quickly to changing market trends and customer demands \\ \hline
        ~ & Minimize stockouts and overstocking & Foster a culture of innovation and continuous improvement at RetailInc \\ \hline
        ~ & Enable data-driven decision-making for inventory management & ~ \\
    \end{tabular}
\end{table}

\subsection{Design Thinking "Hills"}
Table 1, below, provides a record of outputs from ChatGPT Feb 13 Release \cite{ChatGPTReleaseNotes}, prompted with the CHAI-DT framework to perform a "Hills" DT exercise step-by-step \textit{(prompt is available for view in the Appendix)}. 

Hills are "concise statements of the goals we aim to help our users accomplish" \cite{IBM-approach-EDT}, and are built by participants observing and reflecting on a narrow context to individually contribute artifacts (often in the form of sticky notes, one idea written on each) within 3 criteria: 

\begin{itemize}
\item {\textbf{"Who"}} or relevant people within the context we are designing within.
\item{\textbf{"What"}} or enablements provided to the people in our "Who" section. 
\item{\textbf{"Wow"}} or the value differentiator/impact we are providing in our solution. 
\end{itemize}

Typically, after participants write many artifacts within each criteria, the group then clusters the individual artifacts into thematic groups, and reflect on emergent themes to build ideas collaboratively. These clusters are useful in co-creation as they organize highlight ideas from the many perspectives present in the group.

When instructing instructing ChatGPT to perform this exercise all at once, the results worked well. However, when the model is prompted to perform the activity one step at a time, we are able to test its capability to reflect on the context provided, as well as the groups/it's own previous artifacts, to generate \textit{observational} artifacts to the group e.g. users derived from the context, as well as \textit{creative} artifacts such as assumed problems they face and new ideas valuable to them. By allowing the model to reflect on it's own previous outputs, the overall results when instructed to perform the Hills activity step-by-step show higher quality and more relevant artifacts.

While the Table 1 outputs are from a single one-on-one test, they are representative of typical, repeatable patterns seen in multiple tests. Given these results, we can see that ChatGPT was able to generate content relevant to the context of the session with reasonable quality, creativity, and usefulness to cooperative ideation. The models response was also in a manner that follows the instructions provided and can easily be incorporated into a live Design Thinking session.

\subsection{Context}
The {\verb|Context|} sub-prompt provided in this test was a fake scenario generated by ChatGPT Jan 9 Release \cite{ChatGPTReleaseNotes}, solely intended for testing different DT activities: 

\textit{RetailInc, a large retail chain, and IBM have come together to improve their inventory management system. The current system is outdated and often leads to stockouts and overstocking, resulting in lost sales and wasted resources. IBM brings their expertise in data analytics and artificial intelligence to the partnership, while RetailInc provides their extensive data on sales and inventory levels. Together, they aim to create a new system that can accurately predict sales trends and optimize inventory levels in real-time, leading to increased efficiency and profits for the retail chain.}

\subsection{"Self"-acknowledgement and Collaborative Behavior}

Interestingly, after ChatGPT completed each step of the Hills exercise, it also made a point to add disclaimers: 

\textit{"Note: These are just some potential users [or "enablements that the system could provide" or "potential market values or differentiators that the system could provide"], and the team may need to further refine the list based on their research and understanding of the user needs."}. 

This behavior frames the content generated as \textit{intended} for collaboration with others, and in doing so, aligns to Kantosalo's framework by creating an interaction between the Human/s + AI \textit{collective} to \textit{collaborate} through individual \textit{contributions} for the \textit{community} identified in the \textit{context} provided. 

\section{Discussion}
Through the development of the CHAI-DT prompting framework, I have demonstrated that ChatGPT has the capability to provide useful, creative artifacts in live co-creation activities. This method of prompt-engineering is limited in that it is applied to the context of corporate co-creation often utilized for business value creation, however it is adaptable to multiple different styles of co-creative, instruction-based collaboration and further research covering integration into other group-focused co-creative activities is needed. However, the implications for this mode of Human/s + AI co-creation are wide reaching and provide a space for future research into how people or groups of people collaborate, partner, and interact with generative AI agents to enhance collective creativity in other domains and contexts, as well as informs future attempts to design AI agents intended for collaboration, cooperation, and co-creation. 

By introducing a real-world implementation for co-creative generative agents which align to Kantosalo's 5Cs framework, I infer that, through similar or iterative methods, there is opportunity to further explore this space. Engaging methods like CHAI-DT in co-creation, especially when combined with other systems that enable more advanced interactions with Generative AI agents, opens the door to entirely new forms of creative collaboration, as well as environments or methods to enhance inter-disciplinary communication and co-creativity.

Future research is also recommended to address potential risks using generative AI agents in these environments, covered well by Buschek et. al. such as bias, misinformation, conflict of responsibility, and potential exposure to private or inappropriate data \cite{Buschek2021NinePP}, as well as potential for participant distraction, less useful or harmful ideation, misuse of the framework by malicious actors, or potential harm to the communities we are ideating for.

\section{Conclusion}
In this paper, I have introduced CHAI-DT, a novel approach to prompting conversational agents such as ChatGPT to engage in 'Human/s + AI' co-creative efforts, and have explored the potential of integrating creative generative AI agents within live Design Thinking, as well as the implications of this approach and some of the potential risks associated. I also suggest more research be done considering the risks, benefits, and potential future application or evolution of similar frameworks (both useful and harmful) to better understand the future this technology and the framework implies, so we may shape the use of these models in co-creative contexts to be useful, safe, and ultimately enhance the creative capabilities of people and communities.

\bibliographystyle{ACM-Reference-Format}
\bibliography{citations}


\pagebreak

\appendix{Appendix}
\section{CHAI-DT "Hills" Prompt}
We are conducting a "Hills" Design Thinking exercise.

Activity Explanation:
A Hill is a user-centered statements of intent that the entire team can rally around – so that everyone is pulling in the same direction. Hills describe something a specific user is enabled to do, not a specific implementation. They give teams the creative space they need to come to breakthrough ideas, without the need for detailed requirements. Write Hills at the beginning of a project or initiative, after you’ve identified the real needs of your users. Think of it as a team mission statement. The practice of Hills ensures that our entire team is aligned around doing what’s right for the user. They’re about keeping all of our teammates — across disciplines, oceans, and timezones — working in sync on the best experience for the user. 

A hill is a written statement that the team writes and agrees upon together. They’re fairly straightforward with three parts:

The who is very clear and specific on the type of user that we are focusing on, so that could be a tech seller or a business analyst or an HR manager. 

The what is the second part. We’ve referred to these before as a user enablement. This is a specific task, or a specific thing that the user will be able to get done or will be able to accomplish. The what for a hill may talk about being able to deploy or being able to create something or be able to run some sort of analysis. So the what is a specific enablement that the user, mentioned in your who, is able to do. 

And last but not least, a hill includes a wow - who, what, wow. This is a specific market value or differentiating statement. It extends the what into that territory of absolute delight that we know that is required of all of our work with users these days. If you were to read your hill to a sponsor user, which you should, their reaction should be “YES. I want that. I need that. When can I have that?”

So three parts to every hill: a who, a what and a wow.

Example of a good Hill Statement: "Within selected product categories, requestors can find product matches for their search queries using natural, English-language conversation."

Instructions for this activity:
1. Create the list “Who” 
2. Create the list “What”
3. Create the list “Wow.”
4. Diverge on many ideas for each section and quickly share them with your teammates. Build off of others’ ideas, but focus on quantity over quality and avoid drifting into features or talking about implementation details.
5. Build your hill statement(s) using your ideas for “Who,” “What,” and “Wow.” 

Relevant Activity Context:
RetailInc, a large retail chain, and IBM have come together to improve their inventory management system. The current system is outdated and often leads to stockouts and overstocking, resulting in lost sales and wasted resources. IBM brings their expertise in data analytics and artificial intelligence to the partnership, while RetailInc provides their extensive data on sales and inventory levels. Together, they aim to create a new system that can accurately predict sales trends and optimize inventory levels in real-time, leading to increased efficiency and profits for the retail chain.

Given the above context, perform Step 1 of the exercise.

\end{document}